\documentclass[aps,prd,reprint,showpacs,nofootinbib,twoside,notitlepage]{revtex4-1} 
\usepackage{graphicx}
\usepackage{euscript,amssymb}
\usepackage{amsfonts}
\usepackage{amssymb}
\usepackage{amsbsy}
\usepackage{amsmath}
\usepackage{nccmath}   % for de-centering equations environments  i.e.  fleqn environment within which equations envionments are written
\usepackage{color}
\usepackage{bbold}

\usepackage{subcaption} %allows multiple plots in the same figure
\usepackage{mathtools}

\mathtoolsset{showonlyrefs,showmanualtags}

\newcommand{\be}{\begin{equation}}
  \newcommand{\ee}{\end{equation}}
\newcommand{\ben}{\begin{eqnarray*}}
  \newcommand{\een}{\end{eqnarray*}}
\newcommand{\bea}{\begin{eqnarray}}
  \newcommand{\eea}{\end{eqnarray}}
\newcommand{\bdm}{\begin{displaymath}}
  \newcommand{\edm}{\end{displaymath}}
\newcommand{\ba}{\begin{align}}
  \newcommand{\ea}{\end{align}}

\DeclareMathOperator{\atanh}{artanh}
\DeclareMathOperator{\acoth}{arcoth}

\begin{document}

\title{Towards a quantum Oppenheimer-Snyder model} 

\author{Tim Schmitz}

\email{tschmitz@thp.uni-koeln.de}

\affiliation{Institut f\"ur Theoretische Physik, Universit\"{a}t zu
K\"{o}ln, Z\"{u}lpicher Stra\ss e 77, 50937 K\"{o}ln, Germany}

\date{\today}

\begin{abstract}

We present a consistent canonical formulation of the flat Oppenheimer-Snyder model, including the Schwarzschild exterior. The switching between comoving and stationary observer is realized by promoting the coordinate transformation between dust proper time and Schwarzschild-Killing time to a canonical one. This leads to two different forms of the Hamiltonian constraint, both (almost) deparameterizable with regard to one of these times. A preliminary quantization of these constraints reveals a consistent picture for both observers: the singularity is avoided by a bounce.
	
\end{abstract}

%\pacs{04.60.Ds, %Canonical quantization
%	04.70.Dy, %Quantum aspects of black holes, evaporation, thermodynamics
%	04.20.Fy %Canonical formalism, Lagrangians, and variational principles
%}

\maketitle

%%%%%%%%%%%%%%%%%%%%%%%%%%%%%%%%%%%%%%%%%%%%%
%%%%%%%%%%%%%%% INTRODUCTION %%%%%%%%%%%%%%%%
%%%%%%%%%%%%%%%%%%%%%%%%%%%%%%%%%%%%%%%%%%%%%
\section{Introduction}

The Oppenheimer-Snyder (OS) model\footnote{Two years earlier than Oppenheimer and Snyder, B.\ Datt recognized the usefulness of comoving coordinates for solving Einstein equations \cite{DattOS}, and even found a solution for homogeneous dust. He however did not supply any physical interpretation of his solutions.} \cite{OppenheimerSnyder,DattOS} is the prototypical example for gravitational collapse. It has lastingly shaped our understanding of how black holes form by illustrating the need to consider the viewpoint of two different observers, one comoving with the collapsing matter and one stationary outside of it, to arrive at a complete picture of the process. Our goal is to apply this idea to \emph{quantum} gravitational collapse. In this paper we will to this end lay the groundwork for a quantum OS model, complete with the two complementary observers.

A complete theory of quantum gravity is not available as of yet, but it is nevertheless possible to discuss quantizations of e.g.\ specific cosmological models or models for gravitational collapse in different approaches. Due to its interior being homogeneous, the OS model has the advantage that one can quantize it using methods of quantum cosmology. Moreover, since it shares its dynamics with the often quantized Friedmann models, some of the existing literature in quantum cosmology can be applied to it. Explicitly as a collapse model it has only been discussed sparingly so far, e.g.\ in \cite{LundOS,PelegOS} with regard to singularity avoidance and its mass spectrum. Recently it has been investigated in analogy to the hydrogen atom in \cite{CordaOS}.

We have discussed in a previous work a quantization of the Lema\^itre-Tolman-Bondi (LTB) model for inhomogeneous, spherically symmetric dust collapse, implementing unitary evolution from the point of view of the comoving observer \cite{MeLTB}. There we have shown that the classical singularity is avoided by a bounce: instead of fully collapsing to a singularity the matter configuration re-expands. This result straightforwardly carries over to OS collapse. %Here we will expand on \cite{MeLTB} by also explicitly considering the stationary observer.

Bouncing collapse is a common thread in various approaches to quantizing gravitational collapse. It emerged e.g.\ in \cite{FrolovNullShell}, where an effective one-loop action consisting of an Einstein-Hilbert term plus a Weyl-squared term was used to discuss the quantum corrected trajectory of a self-gravitating null dust shell. A null dust shell was also investigated in the context of quantum geometrodynamics \cite{HajicekNullShells,HajicekKieferNullShells}, where a unitarily evolving wave packet initially peaked on the classical collapsing trajectory was shown to bounce. Bouncing collapse has also emerged in loop quantum gravity and cosmology \cite{AshtekarCosmology,RovelliPlanckStars}, and loop-inspired effective semiclassical models \cite{MalafarinaBounce}. 

As universal as the bounce seems to be, there is no consensus on some aspects crucial for the plausibility of this scenario, namely the behavior of the horizon and the lifetime of the temporary black hole-like object.

Concerning the former, proposals include the vanishing of the apparent horizon during the bounce \cite{MalafarinaBounce,BambiBounce,BarceloBounce3,BarceloBounce2} and a transition between black- and white hole horizon via a superposition of the two \cite{HajicekQuantumNullShells,HajicekKieferNullShells}. We have seen in \cite{MeLTB} that the latter is a plausible scenario for the LTB model. The mechanism by which quantum gravitational effects reach the horizon, usually a low curvature region of spacetime, is also a matter of debate. Ideas available are e.g.\ an accumulation of quantum effects over time \cite{HaggardRovelliBounce} or a shockwave going outward from the center of the collapsing body \cite{BarceloBounce2,BarceloBounce3}.

Of greater importance is the lifetime. If it turns out to be too short, bouncing collapse would be immediately ruled out by the fact that we do observe seemingly stationary black holes. In fact, across various models there has emerged the result that the pure transition from collapse to expansion has a lifetime proportional to the mass of the collapsing matter \cite{AmbrusHajicekLifetime,BarceloLifetime,ChristodoulouLifetime,ChristodoulouLifetime2}, which would mean that a solar mass black hole would decay after a few microseconds. The question is now whether one should really take this timescale to be the total lifetime of the black hole. In \cite{ChristodoulouLifetime} it was proposed that this transition timescale should be complemented by a timescale associated to the process taking place should this transition fail, in analogy to an alpha-particle tunneling out of a nucleus. There the tunneling time is also significantly shorter than the total lifetime of the decaying nuclear core. We have followed this proposal in \cite{MeLTB}, leading to a lifetime proportional to the mass of the dust cloud cubed. Another mechanism that could increase the lifetime is the transition from expansion back to collapse due to white hole instabilities \cite{BarceloBounce1}. 

For a more complete review of the above, see \cite{MalafarinaBounceRev}.

It is apparent that there are many open question regarding the scenario of bouncing collapse. We want to try to gain more insight into these issues by  taking the lessons learned from the classical OS model to heart, and consider both the comoving and the stationary observer. In this way we hope to arrive at a more complete picture of bouncing collapse and all it entails. Here we will work towards this goal, building on \cite{MeLTB}. 

We will proceed here as follows. In Sec.\ \ref{ch:chapter_2} we will present a canonical formulation of the complete OS model, with a Schwarzschild exterior. This has previously been attempted in \cite{CasadioOS}, but in our opinion not quite satisfactorily due to the treatment of the new boundary term at the surface of the collapsing dust cloud. For simplicity we will largely restrict ourselves to flat Friedmann models, but we expect the procedure to carry over also to open and closed ones. Using Brown-Kucha\v{r} dust \cite{KucharBrownDust} as the matter content allows us to express the Hamiltonian constraint in a deparameterizable form with regard to dust proper time. We will then demonstrate how one can implement the stationary observer in the canonical formalism by promoting the coordinate transformation between Schwarzschild Killing time and dust proper time to a canonical one. This will lead to a different form of the Hamiltonian constraint that is almost deparameterizable. In Sec.\ \ref{ch:chapter_3} we will then investigate quantization of the constraint in these two forms. First we discuss how our results from \cite{MeLTB} apply to the OS model from the point of view of the comoving observer. Due to the rather unusual structure of the form of the Hamiltonian constraint relevant for the stationary observer, we will only be able to present an heuristic discussion of its quantization. We leave a more rigorous investigation for upcoming work. Finally we conclude in Sec.\ \ref{sec:chapter4}.

We finally want to note that we are also currently investigating a different canonical formulation of the OS model \cite{MeOS}, where in contrast to the present efforts the foliation of spacetime is fixed.

In the following we will use units where $G=c=1$.

%%%%%%%%%%%%%%%%%%%%%%%%%%%%%%%%%%%%%%%%%%%%%%%%%%%%%%%%%%%%%%%%%%

\section{Canonical formalism for the OS model} \label{ch:chapter_2}

\subsection{Partial symmetry reduction}

We will start from the ADM-decomposed action for spherically symmetric gravity with Brown-Kucha\v{r} dust \cite{KucharBrownDust} as matter. Details on how one can derive this action from the Einstein-Hilbert action, as well as the fall-off behavior of the relevant canonical variables, can be found in \cite{KucharSchwarzschild,VazMargLTB1,KieferLTB1}:
\begin{align}
S=&\int dt \int_{0}^{\infty}dr~P_\tau \dot{\tau}+P_R \dot{R}+P_\Lambda\dot{\Lambda}-NH-N^rH_r\nonumber\\
&-\int dt~M_+\dot{T}_+,\label{eq:spherical_action}\\
\end{align}
where
\begin{align}
H =& \frac{\Lambda}{2R^2}~P_\Lambda^2-\frac{1}{R}~P_RP_\Lambda+\frac{RR^{\prime\prime}}{\Lambda}-\frac{\Lambda' R' R}{\Lambda^2}+\frac{{R'}^2}{2\Lambda}\nonumber\\
&-\frac{\Lambda}{2}+P_\tau\sqrt{1+\frac{\tau'^2}{\Lambda^2}},\\
H_r=&P_RR'-P_\Lambda'\Lambda+P_\tau\tau'.
\end{align} 
Therein $\tau$ is the dust proper time, and $\Lambda$ and $R$ the components of the spherically symmetric spatial metric on the leaves of the foliation with label time $t$,
\begin{equation}
d\sigma^2=\Lambda^2(t,r)\,dr^2+R^2(t,r)\,d\Omega^2.\label{eq:spherical_3metric}
\end{equation}
The ADM boundary term in \eqref{eq:spherical_action} contains the ADM mass of the spacetime $M_+(t)$ and the Schwarzschild Killing time at asymptotic infinity $T_+(t)$. Following \cite{KucharSchwarzschild}, this boundary term will play a role in the canonical formalism later on.

So far the action above describes any spherically symmetric spacetime generated by non-rotating, timelike dust, but in the present work we want to restrict ourselves to the OS model: homogeneous dust with a Schwarzschild exterior. To this end we follow \cite{HajicekKijowskiFluid} and work in coordinates adapted to the discontinuity in the matter content. The surface of the dust cloud will always be at coordinate radius $r_S>0$, and outside of it, $r>r_S$, the rest mass density $\rho$ of the dust vanishes. We can implement this with our canonical variables via the vanishing of $P_\tau$, since one can express it through $\rho$ as 
\begin{equation}
P_\tau=4\pi\,\Lambda\, R^2\,\sqrt{1+\frac{\tau'^2}{\Lambda^2}}\,\rho,\label{eq:P_tau}
\end{equation}
where we refer to \cite{KucharBrownDust} for details.

Inside of the dust cloud we want to restrict ourselves to homogeneous dust. It is then natural to also restrict the ADM foliation for $r\leq r_S$ such that its leaves coincide with hypersurfaces of constant $\rho$ and $\tau$. As is well known the corresponding spacetime metric has to be of the form
\begin{equation}
ds^2=-\bar{N}^2(t)\,dt^2+a^2(t) \left(\frac{dr^2}{1-\epsilon r^2}+r^2\,d\Omega^2 \right),\label{eq:homogeneous_4metric}
\end{equation} 
where $\epsilon\in\{\pm1,0\}$ gives the sign of the curvature of constant time slices, and $\bar{N}$ and $a$ are positive. For $\epsilon=+1$ the radial coordinate $r$ is smaller than $1$. Note that for now we leave $\epsilon$ open, later on restricting to $\epsilon=0$. By comparison with the general ADM line element (see e.g.\ \cite{KieferQuantumGravity}) we see that we can implement the restriction of the foliation described above by
\begin{align}
N &= \bar{N},\label{eq:falloff_N}\\
N^r &= 0, \label{eq:falloff_Nr}
\end{align} 
for $r\leq r_S$.

Furthermore, we can fix the behavior of the canonical variables $\Lambda$ and $R$ for $r\leq r_S$ as
\begin{align}
\Lambda &= \frac{a}{\sqrt{1-\epsilon r^2}},\label{eq:falloff_Lambda}\\
R &= a\,r,\label{eq:falloff_R}
\end{align} 
by comparing \eqref{eq:spherical_3metric} with \eqref{eq:homogeneous_4metric}. Consequently we can also note that $P_\tau$ behaves as
\begin{equation}
P_\tau=\frac{4\pi r^2\,a^3}{\sqrt{1-\epsilon r^2}}\,\rho\equiv\frac{r^2}{V_S\sqrt{1-\epsilon r^2}}\,\bar{P}_\tau,
\end{equation}
with
\begin{equation}
	V_S=\int_{0}^{r_S}dr~\frac{r^2}{\sqrt{1-\epsilon r^2}}.
\end{equation}
The significance of $\bar{P}_\tau$, which only depends on time, will be apparent shortly. Lastly we also express $P_\Lambda$ and $P_R$ in terms of the scale factor and its time derivative as
\begin{align}
P_\Lambda &= -\frac{R}{N}\left( \dot{R}-R'N^r\right)=-\frac{a\dot{a}\, r^2}{\bar{N}}\label{eq:P_Lambda} ,\\
P_R &= -\frac{\Lambda}{N}\left( \dot{R}-R'N^r\right) -\frac{R}{N}\left( \dot{\Lambda}-\left(\Lambda N^r\right) '\right) \nonumber\\
 &=-\frac{2\,a\dot{a}\, r}{\bar{N}\sqrt{1-\epsilon r^2}}\label{eq:P_R} . 
\end{align}

Now we are in a position to replace the canonical coordinates $\Lambda$ and $R$ with the scale factor $a$ in the region $r\leq r_S$, we just need to find a canonical momentum for it. To this end we consider the Liouville form from \eqref{eq:spherical_action} restricted to $r\leq r_S$,
\begin{align}
\int_{0}^{r_S}dr~P_\tau &\dot{\tau}+P_R \dot{R}+P_\Lambda\dot{\Lambda}\\ =& \int_{0}^{r_S}dr\left( \frac{r^2}{V_S\sqrt{1-\epsilon r^2}}\,\bar{P}_\tau \dot{\tau}
-\frac{3\,a\dot{a}^2\, r^2}{\bar{N}\sqrt{1-\epsilon r^2}}\right) \\
=& \bar{P}_\tau\dot{\tau}-\frac{3V_S\,a\dot{a}}{\bar{N}}\,\dot{a}.
\end{align}
This allows us to identify $\bar{P}_\tau$ as the momentum to $\tau$, and as the momentum canonically conjugate to $a$
\begin{equation}
p=-\frac{3V_S\,a\dot{a}}{\bar{N}},
\end{equation}
from which with the help of \eqref{eq:P_Lambda} and \eqref{eq:P_R} directly follows that
\begin{align}
P_\Lambda &= \frac{r^2}{3V_S}~p,\label{eq:falloff_PLambda}\\
P_R & = \frac{2r}{3V_S\sqrt{1-\epsilon r^2}}~p. \label{eq:falloff_PR}
\end{align}

Finally we can compute the Hamiltonian for $r\leq r_S$:
\begin{align}
\int_{0}^{r_S}dr~ NH+N^rH_r %=&\int_{0}^{r_S}dr~\frac{r^2}{V_S\sqrt{1-\epsilon r^2}}\,\bar{N}\\&\times\left( -\frac{p^2}{6V_S\,a}-\frac{3V_S}{2}\epsilon a +\bar{P}_\tau \right)\\ 
=&\,\bar{N}\left( -\frac{p^2}{6V_S\,a}-\frac{3V_S}{2}\epsilon a +\bar{P}_\tau \right).
\end{align}
The total action we have achieved by partial symmetry reduction of the phase space is then
\begin{align}
S=&\int dt~\left[  p\dot{a}+\bar{P}_\tau\dot{\tau}-\bar{N}\bar{H}\phantom{\int}\right. \\
 +&\left.\int_{r_S}^{\infty}dr\,\left( P_R \dot{R}+P_\Lambda\dot{\Lambda}-NH-N^rH_r\right)-M_+\dot{T}_+\right] , \label{eq:symmetry_reduced_action}\\
H =& \frac{\Lambda}{2R^2}~P_\Lambda^2-\frac{1}{R}~P_RP_\Lambda+\frac{RR^{\prime\prime}}{\Lambda}-\frac{\Lambda' R' R}{\Lambda^2}+\frac{{R'}^2}{2\Lambda}-\frac{\Lambda}{2},\\
H_r=&P_RR'-P_\Lambda'\Lambda,\\
\bar{H}=&-\frac{p^2}{6V_S\,a}-\frac{3V_S}{2}\epsilon a +\bar{P}_\tau.
\end{align}

Note that the newly gained Hamiltonian constraint for the inside of the dust cloud is equivalent to the one of a Friedmann model with Brown-Kucha\v{r} dust and a vanishing cosmological constant, see \cite{MaedaFriedmann}. In fact, we could have simply written down the total action \eqref{eq:symmetry_reduced_action} by adding up the actions for the Friedmann model and Schwarzschild in the exterior. Our more elaborate procedure additionally gives us the fall-off behavior of the canonical variables of the exterior when approaching $r\to r_S$: imposing that the canonical variables and those derivatives of them we will encounter behave according to \eqref{eq:falloff_N}, \eqref{eq:falloff_Nr}, \eqref{eq:falloff_Lambda}, \eqref{eq:falloff_R}, \eqref{eq:falloff_PLambda}, and \eqref{eq:falloff_PR} for $r\to r_S$ ensures a matching between interior and exterior that is as smooth as needed.

Employing this fall-off behavior one can see that the Hamiltonian constraint has a discontinuity at the surface of the star: $H$ approaches $\bar{H}-\bar{P}_\tau$ for $r\to r_S$, but on the cloud's surface the constraint is given by just $\bar{H}$. It is thus important to note that we have to explicitly exclude the surface of the dust cloud, $r=r_S$, from the exterior. Note that the momentum constraint is continuous, $H_r$ vanishes for $r\to r_S$.

From this also follows that the total Hamiltonian, $NH+N^rH_r$ in the exterior and $\bar{N}\bar{H}$ in the interior, has the same discontinuity corresponding to the matter content of the cloud.

Apart from the implementation of matter, our results so far are close to those of \cite{CasadioOS}, despite proceeding along slightly different lines. It is in the following that the two treatments will diverge significantly.

\subsection{Kucha\v{r}'s canonical transformations}

In order to simplify the constraints in the exterior we follow \cite{KucharSchwarzschild} and perform two canonical transformations in succession. The first one replaces the metric component $\Lambda$ by the mass of the dust cloud $M$, and the second will in turn replace $M$ by the Schwarzschild Killing time $T$. Special attention has to be paid to boundary terms arising from these transformations, since in contrast to \cite{KucharSchwarzschild,VazMargLTB1,KieferLTB1} one of the boundaries is at finite coordinate radius $r_S$. These terms could directly influence the dust cloud's interior, since the corresponding part of the action is essentially also a boundary term.

Consider first the transformation $(\Lambda,R,P_\Lambda,P_R)$ to $(M,\mathsf{R},P_M,P_\mathsf{R})$ according to
\begin{align}
M =& \frac{R}{2}(1-F),\\
\mathsf{R} =& R,\\
P_M =& \frac{\Lambda P_\Lambda}{R F},\\
P_\mathsf{R} =& P_R - \frac{\Lambda P_\Lambda}{2 R} -\frac{\Lambda P_\Lambda}{2 R F}\\
& - \frac{1}{R \Lambda^2 F}\left[ \left(\Lambda P_\Lambda \right)'RR'- \left(RR' \right)'\Lambda P_\Lambda \right],
\end{align}
where
\begin{equation}
F = \frac{R'^2}{\Lambda^2} - \frac{P_\Lambda^2}{R^2}.
\end{equation}\
This transformation generates a boundary term vanishing at $r\to\infty$, as detailed in \cite{KucharSchwarzschild}, but not at $r\to r_S$,
\begin{align}
\int_{r_S}^{\infty}dr~P_R\dot{R}+P_\Lambda\dot{\Lambda} &-\int_{r_S}^{\infty}dr~P_\mathsf{\mathsf{R}}\dot{\mathsf{R}}+P_M\dot{M} \\&= \left. \frac{R\dot{R}}{2}\,\ln\left| \frac{RR'-\Lambda P_\Lambda}{RR'-\Lambda P_\Lambda}\right| \right|_{r\to r_S}\\
&= \frac{a\dot{a}\,r_S^2}{2}\,\ln\left| \frac{a-\frac{r_S}{3V_S\sqrt{1-\epsilon r_S^2}}\,p}{a+\frac{r_S}{3V_S\sqrt{1-\epsilon r_S^2}}\,p}\right|.
\end{align}
We will postpone the discussion about how to deal with this additional term  for a little while until after the second canonical transformation. For now we just want to comment on the behavior of the new canonical variable $M$ at the surface of the star,
\begin{align}
M(r_S)\equiv \bar{M} &= \frac{r_S^3}{3V_S}\left(\frac{p^2}{6V_S\,a}+\frac{3V_S}{2}\epsilon a \right)\\
& =  \frac{r_S^3}{3V_S}\left(\bar{P}_\tau - \bar{H}\right). \label{eq:mass_on_surface}
\end{align}
We can thus express $\bar{M}$ as
\begin{equation}
\bar{M}\approx\frac{r_S^3}{3V_S}\bar{P}_\tau= \frac{4\pi}{3}r_S^3\,a^3\,\rho,
\end{equation}
where we once again replaced $\bar{P}_\tau$ by the mass density $\rho$. We denote by `$\approx$' a weak inequality, valid when the constraints vanish. The mass of the Schwarzschild exterior corresponds to the total mass of the interior. Appropriate matching conditions give the same result, see e.g.\ \cite{PoissonRelativistsToolkit}. This serves as a consistency check for our canonical treatment so far.

Now we implement the second canonical transformation $(M,P_M)$ to $(T,P_T)$. The crucial observation is that on-shell one can identify $P_M=-T'$ (see \cite{KucharSchwarzschild} for details), so we will define
\begin{equation}
T = T_++\int_{r}^{\infty}dr'~P_M \label{eq:defT}
\end{equation}
and consider the relevant terms in the Liouville form to find the corresponding momentum:
\begin{align}
-M_+\dot{T}_++\int_{r_S}^{\infty}dr~P_M\dot{M}=&-M_+\dot{T}_+-\int_{r_S}^{\infty}dr~T'\dot{M}\\
=& - \bar{M} \dot{\bar{T}} - \int_{r_S}^{\infty}dr~M'\dot{T}\\
+\frac{d}{dt}\left(\bar{M} \bar{T}\phantom{\int}\hspace*{-1.1em}\right. &\left.- M_+  T_+ +\int_{r_S}^{\infty}dr~M'T \right),	
\end{align}
where $\bar{T}=T(r_S)$. The total time derivative can be discarded. We can immediately identify $P_T=-M'$, and the additional boundary term $-\bar{M}\dot{\bar{T}}$. Kucha\v{r} has shown in \cite{KucharSchwarzschild} that the constraint system $H=0= H_r$ is equivalent to the far simpler system
\begin{equation}
	P_\mathsf{R}=0= P_T. \label{eq:new_constraints_ext}
\end{equation}

The behavior of $P_T$ and $P_\mathsf{R}$ when approaching the surface of the dust cloud,
\begin{align}
P_T(r\to r_S) =& -\frac{r_S^2}{V_S}\left(\frac{p^2}{6V_S\,a}+\frac{3V_S}{2}\epsilon a \right)=\frac{r_S^2}{V_S}(\bar{H}-\bar{P}_\tau),\\
P_\mathsf{R}(r\to r_S) =& \frac{r_S\,p}{2V_S\sqrt{1-\epsilon r_S^2}} \left(1-\frac{1}{1-\frac{2\bar{M}}{ar_S}} \right)\\
 =& \frac{r_S^4}{3V_S^2\sqrt{1-\epsilon r_S^2}} \frac{p}{ar_S-2\bar{M}}(\bar{H}-\bar{P}_\tau),
\end{align}
shows that the new constraints have the same discontinuity at the surface of the dust cloud as the old ones.

Keeping track of the boundary terms and using the new constraint system, our action now takes the form
\begin{multline}
S=\int dt ~ p\dot{a}+\bar{P}_\tau\dot{\tau}\\+\frac{a\dot{a}\,r_S^2}{2}\,\ln\left| \frac{a-\frac{r_S}{3V_S\sqrt{1-\epsilon r_S^2}}\,p}{a+\frac{r_S}{3V_S\sqrt{1-\epsilon r_S^2}}\,p}\right| - \bar{M} \dot{\bar{T}}-\bar{N}\bar{H}\\+\int dt\int_{r_S}^{\infty}dr~\left( P_\mathsf{R} \dot{\mathsf{R}}+P_T\dot{T}-N^\mathsf{R}P_\mathsf{R}-N^TP_T\right) , \label{eq:action_1}
\end{multline}
where the Lagrange multipliers were redefined into $N^R$ and $N^T$ as needed. It is apparent that the action for the interior of the dust cloud is not canonical anymore due to the additional boundary terms. In order to remedy this problem we note that $\bar{T}$ does not directly follow from \eqref{eq:defT} and the behavior of $P_M$ for $r\to r_S$. We thus have to choose it such that the additional terms in the Liouville form disappear, making the preceding series of transformations truly canonical.

We will for simplicity restrict ourselves to flat Friedmann models in the interior, $\epsilon=0$. $V_S$ then simply is $\frac{1}{3}r_S$, and it will be convenient to rescale our canonical variables according to $\bar{R}=a\,r_S$ and $\bar{P}_R=p/r_S$. The new canonical variable $\bar{R}$ is then the physical radius of the dust cloud. We then consider the unwanted terms in \eqref{eq:action_1}, inserting \eqref{eq:mass_on_surface}:
\begin{widetext}
\begin{align}
\frac{1}{2}\,\bar{R}\dot{\bar{R}}\,\ln\left| \frac{\bar{R}-\bar{P}_R}{\bar{R}+\bar{P}_R}\right| - \frac{\bar{P}_R^2}{2\bar{R}} \dot{\bar{T}} &= \frac{1}{4}\frac{\partial \bar{R}^2}{\partial t}\,\ln\left| \frac{\bar{R}-\bar{P}_R}{\bar{R}+\bar{P}_R}\right| - \frac{\bar{P}_R^2}{2} \, \frac{\partial}{\partial t}\left(\frac{\bar{T}}{\bar{R}} \right) -\frac{\bar{P}_R^2\bar{T}}{4\bar{R}^3}\,\frac{\partial \bar{R}^2}{\partial t}\\
&= \frac{\bar{R}^2}{4}\frac{\partial}{\partial t}\,\ln\left| \frac{\bar{R}+\bar{P}_R}{\bar{R}-\bar{P}_R}\right| - \frac{\bar{P}_R^2}{4}\, \frac{\partial}{\partial t}\left(\frac{\bar{T}}{\bar{R}} \right) +\frac{\bar{R} \bar{T}}{4}\,\frac{\partial }{\partial t}\left( \frac{\bar{P}_R^2}{\bar{R}^2}\right) + \dot{K}(\bar{P}_R,\bar{R})\\
&=\frac{\bar{P}_R^4}{4\bar{R}^2}\left(\frac{\bar{R}^4}{\bar{P}_R^4}\,\frac{\partial}{\partial t}\,\ln\left| \frac{\bar{R}+\bar{P}_R}{\bar{R}-\bar{P}_R}\right|-\frac{\partial}{\partial t}\left(\frac{\bar{R} \bar{T}}{\bar{P}_R^2}\right) \right) + \dot{K}(\bar{P}_R,\bar{R})\\
&=\frac{\bar{P}_R^4}{4\bar{R}^2}\,\frac{\partial}{\partial t}\left(\ln\left| \frac{\bar{R}+\bar{P}_R}{\bar{R}-\bar{P}_R}\right| - \frac{2\bar{R}^3}{3\bar{P}_R^3} - \frac{2\bar{R}}{\bar{P}_R} -\frac{\bar{R} \bar{T}}{\bar{P}_R^2}\right) + \dot{K}(\bar{P}_R,\bar{R}).
\end{align}
\end{widetext}
The exact form of the function $K$ is not relevant here. We can now directly read off that setting
\begin{equation}
\bar{T} =  - \frac{2\bar{R}^2}{3\bar{P}_R} - 2\bar{P}_R +\frac{\bar{P}_R^2}{\bar{R}}\,\ln\left| \frac{\bar{R}+\bar{P}_R}{\bar{R}-\bar{P}_R}\right| + A\,\frac{\bar{P}_R^2}{\bar{R}},\label{eq:T_S_canonical}
\end{equation}
where $A$ is some undetermined constant, will bring the action \eqref{eq:action_1} for $\epsilon=0$ into canonical form,
\begin{align}
S=&\int dt~ \bar{P}_R\dot{\bar{R}}+\bar{P}_\tau\dot{\tau}-\bar{N}\bar{H}\\
+&\int_{r_S}^{\infty}dr~\left( P_\mathsf{R} \dot{\mathsf{R}}+P_T\dot{T}-N^\mathsf{R}P_\mathsf{R}-N^TP_T\right), \label{eq:action_2}\\
\bar{H}=&-\frac{\bar{P}_R^2}{2\bar{R}}+\bar{P}_\tau.\label{eq:hamiltonian_interior}
\end{align}
We will shortly show that \eqref{eq:T_S_canonical} is consistent with the interpretation of $T$ as the Schwarzschild Killing time.

It can easily be shown that this action indeed describes spatially flat OS dust collapse. The action for the exterior tells us that there the leaves of the foliation are indeed embedded in a Schwarzschild spacetime via the variables $\mathsf{R}$ and $T$, and its dynamics are independent of this embedding, completely analogous to \cite{KucharSchwarzschild}. The mass of this Schwarzschild exterior is given by the dust cloud: as a constraint, $P_T=-M'\approx0$, so the mass $M$ is determined by its value on the surface of the dust cloud, $\bar{M}=\frac{\bar{P}^2}{2\bar{R}}\approx\bar{P}_\tau$.

The dynamics of the interior of the dust cloud are generated by the Hamiltonian $\bar{N} \bar{H}$:
\begin{align}
\dot{\bar{R}}&=-\bar{N}\,\frac{\bar{P}_R}{\bar{R}},\\
\dot{\bar{P}}_R&=-\bar{N}\,\frac{\bar{P}_R^2}{2\bar{R}^2},\\
\dot{\tau}&=\bar{N},\\
\dot{\bar{P}}_\tau&=0.
\end{align}

Using the constraint $\bar{H}=0$ allows us to easily solve the equations of motion by using $\tau$ as a time parameter,
\begin{equation}
\bar{P}_\tau\approx\frac{\bar{P}_R^2}{2\bar{R}}=\frac{\bar{R}\dot{\bar{R}}^2}{2\dot{\tau}^2}=\frac{\bar{R}}{2}\left(\frac{\partial \bar{R}}{\partial \tau} \right)^2.
\end{equation}
Expressing $\bar{P}_\tau$ in terms of $\rho$ shows that this is simply the equation of motion for a flat Friedmann model with dust as matter. Solutions are
\begin{align}
\bar{R}^\frac{3}{2}(\tau)&=\pm\frac{3}{2}\sqrt{2\bar{P}_\tau}\,\tau+\bar{R}^\frac{3}{2}(0),\\
\bar{P}_R(\tau)&=-\bar{R}(\tau)\frac{\partial \bar{R}}{\partial \tau}=\mp\sqrt{2\bar{P}_\tau\, \bar{R}(\tau)}.\label{eq:eom1}
\end{align}
Plugging these into \eqref{eq:T_S_canonical} we find
\begin{equation}
\bar{T} =\tau + B \pm 2\sqrt{2\bar{P}_\tau}\left[ \sqrt{\bar{R}} - \sqrt{\frac{\bar{P}_\tau}{2}}\,\ln\left| \frac{\sqrt{\bar{R}}+ \sqrt{2\bar{P}_\tau}}{\sqrt{\bar{R}}- \sqrt{2\bar{P}_\tau}}\right|\right],\label{eq:T_S_coordinate}
\end{equation}
where $B=2\,A\,\bar{P}_\tau\pm\frac{2\bar{R}^\frac{3}{2}(0)}{3\sqrt{2\bar{P}_\tau}}$ is constant. We see that on-shell and for $\bar{R}>2\bar{P}_\tau$, apart from an irrelevant constant shift, $\bar{T}$ indeed matches the Schwarzschild Killing time on the surface of the dust cloud, in its form known from the transformation between Schwarzschild- and Painlev\'{e}-Gullstrand coordinates, see e.g. \cite{KieferLTB1}. The different signs in it correspond to an expanding and a collapsing dust cloud, respectively. The expression above is furthermore even well defined inside the horizon, due to the absolute value in the logarithm. This will play a role during quantization.

We expect this procedure to also work for $\epsilon=\pm1$. Instead of Painlev\'{e}-Gullstrand time one would then find for $\tau(\bar{T})$ different time coordinates adapted to the comoving observer in each case.

\subsection{Switching between observers as a canonical transformation}

We now want to introduce Schwarzschild Killing time into the phase space. To this end we can use the prescription \eqref{eq:T_S_coordinate} (with $B=0$) to introduce $\bar{T}$ as a canonical variable, promoting the transformation from Schwarzschild- to Painlev\'{e}-Gullstrand coordinates to a canonical transformation on phase space.

We consider a generating function of the third kind $\Omega(\bar{P}_R,\bar{P}_\tau;\bar{\mathsf{R}},\bar{T})$, where $\bar{\mathsf{R}}=\bar{R}$. $\Omega$ can then be determined from
\begin{align}
\bar{R}&=-\frac{\partial \Omega}{\partial \bar{P}_R}\overset{!}{=}\bar{\mathsf{R}},\\
\tau&=-\frac{\partial \Omega}{\partial \bar{P}_\tau}\\&\overset{!}{=}\bar{T}\mp 2\sqrt{2\bar{P}_\tau}\left[ \sqrt{\bar{\mathsf{R}}} - \sqrt{\frac{\bar{P}_\tau}{2}}\,\ln\left| \frac{\sqrt{\bar{\mathsf{R}}}+ \sqrt{2\bar{P}_\tau}}{\sqrt{\bar{\mathsf{R}}}- \sqrt{2\bar{P}_\tau}}\right|\right],
\end{align}
to be of the form
\begin{multline}
\Omega=-\bar{P}_R\bar{\mathsf{R}}-\bar{P}_\tau \bar{T}\\ \pm \int d\bar{P}_\tau~2\sqrt{2\bar{P}_\tau}\left[ \sqrt{\bar{\mathsf{R}}} - \sqrt{\frac{\bar{P}_\tau}{2}}\,\ln\left| \frac{\sqrt{\bar{\mathsf{R}}}+ \sqrt{2\bar{P}_\tau}}{\sqrt{\bar{\mathsf{R}}}- \sqrt{2\bar{P}_\tau}}\right|\right]\\ +\Theta(\bar{T},\bar{\mathsf{R}}),
\end{multline}
where $\Theta$ is an undetermined function. Note that we have to restrict $\bar{P}_\tau$ to be positive. We will see that the final form of $\bar{H}$ can easily be extended to again include negative $\bar{P}_\tau$.

Now we can determine the momenta conjugate to $\bar{\mathsf{R}}$ and $\bar{T}$ according to
\begin{alignat}{3}
\bar{P}_T&=-\frac{\partial \Omega}{\partial \bar{T}}&&=\bar{P}_\tau-\frac{\partial \Theta}{\partial \bar{T}},\\
\bar{P}_\mathsf{R}&=-\frac{\partial \Omega}{\partial \bar{\mathsf{R}}}&&=%\bar{P}_R-\frac{\partial \Theta}{\partial \bar{\mathsf{R}}}\\&~\mp\int d\bar{P}_\tau\,\left[ \sqrt{\frac{2\bar{P}_\tau}{\bar{\mathsf{R}}}} - 2\bar{P}_\tau\,\frac{\partial}{\partial \bar{\mathsf{R}}}\ln\left| \frac{\sqrt{\bar{\mathsf{R}}}+ \sqrt{2\bar{P}_\tau}}{\sqrt{\bar{\mathsf{R}}}- \sqrt{2\bar{P}_\tau}}\right|\right] \\
\bar{P}_R-\frac{\partial \Theta}{\partial \bar{\mathsf{R}}}\mp\int d\bar{P}_\tau~\sqrt{\frac{2\bar{P}_\tau}{\bar{\mathsf{R}}}} \frac{\bar{\mathsf{R}}}{\bar{\mathsf{R}}-2\bar{P}_\tau} \\
&&&=\bar{P}_R-\frac{\partial \Theta}{\partial \bar{\mathsf{R}}}\pm\sqrt{2\bar{P}_\tau\bar{\mathsf{R}}}\\&&&~\pm\bar{\mathsf{R}}\begin{cases}
\atanh\sqrt{\frac{2\bar{P}_\tau}{\bar{\mathsf{R}}}},~&\bar{\mathsf{R}}>2\bar{P}_\tau\\
\acoth\sqrt{\frac{2\bar{P}_\tau}{\bar{\mathsf{R}}}},~&\bar{\mathsf{R}}<2\bar{P}_\tau
\end{cases}.
\end{alignat}
Furthermore we choose $\Theta=0$, such that the momentum $\bar{P}_T=\bar{P}_\tau$ continues to have a nice physical interpretation in terms of the dust cloud's mass.

Due to the different signs in \eqref{eq:T_S_coordinate}, the system has to be split into two distinct parts: one describing collapsing and one expanding dust clouds. In addition, the new momentum $\bar{P}_\mathsf{R}$ diverges at $\bar{\mathsf{R}}=2\bar{P}_\tau$. 

This divergence leads to a disjointed constraint surface, split at the horizon, as can be seen by simplifying the Hamiltonian constraint \eqref{eq:hamiltonian_interior} in terms of the new canonical variables. To this end we first note that
\begin{equation}
\bar{P}_R\pm\sqrt{2\bar{R}\bar{P}_\tau}\approx 0
\end{equation}
is equivalent to \eqref{eq:hamiltonian_interior}, where the signs were chosen to align with the equation of motion \eqref{eq:eom1}; just as in \eqref{eq:T_S_coordinate} the upper sign corresponds to expansion and the lower one to collapse. Inserting the new canonical variables we see that
\begin{equation}
\frac{\bar{P}_\mathsf{R}}{\bar{\mathsf{R}}}\approx\pm\begin{cases}
\atanh\sqrt{\frac{2\bar{P}_T}{\bar{\mathsf{R}}}},~&\bar{\mathsf{R}}>2\bar{P}_T\\
\acoth\sqrt{\frac{2\bar{P}_T}{\bar{\mathsf{R}}}},~&\bar{\mathsf{R}}<2\bar{P}_T
\end{cases}.
\end{equation}
Solving for $\bar{P}_T$ then leads to the final form of the Hamiltonian constraint,
\begin{equation}
\bar{H}_T = \bar{P}_T - \frac{\bar{\mathsf{R}}}{2}\begin{cases}
\tanh^2\frac{\bar{P}_\mathsf{R}}{\bar{\mathsf{R}}},~&\bar{\mathsf{R}}>2\bar{P}_T\\
\coth^2\frac{\bar{P}_\mathsf{R}}{\bar{\mathsf{R}}},~&\bar{\mathsf{R}}<2\bar{P}_T
\end{cases}.\label{eq:hamiltonian_interior2}
\end{equation}
As is apparent, the split between collapse and expansion has disappeared, but there is one at the horizon: passing from $\bar{\mathsf{R}}>2\bar{P}_T$ to $\bar{\mathsf{R}}<2\bar{P}_T$ on the constraint surface is only possible by taking $|\bar{P}_\mathsf{R}|\to\infty$. Since the position of this split still depends on $\bar{P}_T$, we call this constraint \emph{almost} deparameterizable.

%%%%%%%%%%%%%%%%%%%%%%%%%%%%%%%%%%%%%%%%%%%%%%%%%%%%%%%%%%%%%%%%%%%%%%%%%

\section{Quantization} \label{ch:chapter_3}

\subsection{The comoving observer} \label{sec:quant_proper_time}

We first consider the action in the form \eqref{eq:action_2}. Following Dirac's prescription for quantizing constrained systems and using the Schr\"odinger representation leads to the equations
\begin{equation}
\frac{\delta \psi}{\delta \mathsf{R}(r)}=0,\quad\frac{\delta \psi}{\delta T(r)}=0
\end{equation}
corresponding to \eqref{eq:new_constraints_ext} and
\begin{equation}
i\hbar\frac{\partial \psi}{\partial \tau}=\frac{\hbar^2}{2}\bar{R}^{-1+a+b}\frac{\partial}{\partial \bar{R}}\bar{R}^{-a}\frac{\partial}{\partial \bar{R}}\bar{R}^{-b}\psi,\label{eq:proper_time_Seq}
\end{equation}
corresponding to \eqref{eq:hamiltonian_interior} for the wave functional $\psi(\tau,\bar{R};\mathsf{R},T]$, where $a,b\in\mathbb{R}$ control the factor ordering. The first two equations turn the wave functional into a wave function, only depending on the interior degrees of freedom. The last equation is effectively a Schr\"odinger equation with dust proper time as the time parameter. 

This Schr\"odinger equation we have discussed in \cite{MeLTB} for Lema\^{i}tre-Tolman-Bondi collapse. We will briefly recapitulate the results of this previous work and how they relate to the OS model. 

We can make use of the structure of \eqref{eq:proper_time_Seq} and effectively deparameterize the theory, treating $\tau$ as an external parameter. This gives us access to the full structure of quantum mechanics we employed in \cite{MeLTB}: it allowed us to define a Hilbert space, and make the effective Hamiltonian into a self-adjoint operator. The system thus evolves unitarily with dust proper time.

Analysis of the asymptotic behavior of wave packets near the classical singularity $\bar{R}=0$ then showed that the probability distribution for $\bar{R}$ based on these wave packets always vanishes at the singularity, provided the factor ordering fulfills $|1+a|\geq3$ or $|1+a|<2$, regardless of the specific wave packet, or the self-adjoint extension of the Hamiltonian. Furthermore one can choose a self-adjoint extension such that the same is true regardless of the factor ordering. 

Since one expects wave packets to exhibit semiclassical behavior, following the classical trajectories far away from the singularity, we interpreted this as fairly generic avoidance of the singularity by dust collapse when quantum gravity effects are taken into account. This carries over as is to the OS case.

Considering one specific wave packet, we showed that it avoided the singularity via a bounce, transitioning from the classical collapsing trajectory to the expanding one shortly before hitting the singularity. The minimal radius of the dust cloud is then inversely proportional to $M^\frac{1}{3}$. This has the effect that for astrophysically relevant masses this minimal radius is sub-Planckian, for e.g. solar mass the radius is of the order $10^{-13}\,l_P$.

Because we considered an inhomogeneous model for dust collapse in \cite{MeLTB}, another effect emerged: near the singularity, the dust shells could possibly reverse their order. This would lead to a higher minimal radius of the full inhomogeneous dust cloud, since an inner dust shell, with less mass contained inside it, is now the outermost one. This effect can obviously not play a role here, since we have restricted ourselves to homogeneous collapse. 

If our treatment of the inhomogeneous model in \cite{MeLTB} is valid, then quantum cosmology as embedded in a full theory of quantum gravity would be unstable. Perhaps unsurprisingly, homogeneity would necessarily break near the classical singularity. This possibility has been discussed before, e.g. in \cite{KucharRyanInstability} using anisotropic cosmological models. To see it here explicitly suggests that it might also be possible to investigate this phenomenon closer using the OS model. We leave a detailed discussion of this for future work.

%Lastly we have discussed in \cite{MeLTB} a few aspects of the quantum corrected spacetime for inhomogengeous dust collapse suggested by the quantum dynamics. Of greatest importance here are the behavior of the horizon and the lifetime of the temporary black hole. We were able to extract some information concerning these aspects from the quantum model, but reaching definite results was hindered by the limits of this model, chiefly the restriction to dust proper time and with that the point of view of the comoving observer. Here we want to work towards remedying this, by also switching to Killing time and consider the exterior stationary observer.

\subsection{The stationary observer}

We now want to discuss quantization of the system after introducing Schwarzschild Killing time as a canonical variable. The exterior constraints act as in the last subsection, such that the wave function only depends on the interior degrees of freedom. In the interior we have the Hamiltonian constraint \eqref{eq:hamiltonian_interior2}. 

Quantization of \eqref{eq:hamiltonian_interior2} in this form is quite challenging due to the complicated dependency on the momentum, and because it is only almost deparameterizable. We are currently investigating this Hamiltonian in more detail in a phase space approach called coherent state quantization. Here we want to present a preliminary, heuristic way of finding quantum corrected dynamics of this system.

First we want to note that, because $\tanh^2(x)<1$ and $\coth^2(x)>1$, the constraint surface described by \eqref{eq:hamiltonian_interior2} can also be given in terms of
\begin{equation}
\bar{P}_T\approx \frac{\bar{\mathsf{R}}}{2} \tanh^2\frac{\bar{P}_\mathsf{R}}{\bar{\mathsf{R}}}\quad\text{or}\quad\bar{P}_T\approx \frac{\bar{\mathsf{R}}}{2} \coth^2\frac{\bar{P}_\mathsf{R}}{\bar{\mathsf{R}}}.
\end{equation}
A point in phase space is on the constraint surface if either condition is fulfilled. We will hence quantize both partial constraints separately, and will accept as physical states arbitrary linear combinations of solutions to either constraint equation. Note that in this form the constraints are again truly deparameterizable.

Next we want to perform another canonical transformation $(\bar{\mathsf{R}},\bar{P}_\mathsf{R})$ to $(\Phi,\Pi)$, given by
\begin{align}
\Phi &= \frac{\bar{\mathsf{R}}^2}{2}\,\cosh^2\frac{\bar{P}_\mathsf{R}}{\bar{\mathsf{R}}},\\
\Pi &= \tanh\frac{\bar{P}_\mathsf{R}}{\bar{\mathsf{R}}}.
\end{align}
Note that the new momentum is bounded, $|\Pi|<1$. We then replace the constraints by equivalent ones by squaring both sides, and express them in the new variables:
\begin{equation}
\bar{P}_T^2\approx \frac{\Phi}{2} (1-\Pi^2)\Pi^4\quad\text{or}\quad\bar{P}_T^2\approx \frac{\Phi}{2} (1-\Pi^2)\frac{1}{\Pi^4}.
\end{equation}

Canonically quantizing these constraints then gives in momentum representation with regard to $(\Phi,\Pi)$, choosing the most straightforward factor ordering,
\begin{align}
-\hbar^2\frac{\partial^2}{\partial \bar{T}^2}\psi_+(\bar{T},\Pi)-\frac{i\hbar}{2} (1-\Pi^2)\Pi^4 \frac{\partial}{\partial \Pi}\psi_+(\bar{T},\Pi)&=0,\\
-\hbar^2\frac{\partial^2}{\partial \bar{T}^2}\psi_-(\bar{T},\Pi)-\frac{i\hbar}{2} (1-\Pi^2)\frac{1}{\Pi^4} \frac{\partial}{\partial \Pi}\psi_-(\bar{T},\Pi)&=0.
\end{align}
Physical states are then given by
\begin{equation}
\psi(\bar{T},\Pi)=C_+\psi_+(\bar{T},\Pi)+C_-\psi_-(\bar{T},\Pi),
\end{equation}
where $C_\pm$ are arbitrary constants.

Stationary modes are
\begin{equation}
\psi_\pm^\omega(\bar{T},\Pi)=\exp\left[ \frac{i}{\hbar}\omega \bar{T} - \frac{2i}{\hbar}\omega^2f_\pm(\Pi)  \right] ,
\end{equation}
where
\begin{align}
f_+(\Pi)&=-\frac{1}{3\Pi^3}-\frac{1}{\Pi}+\frac{1}{2}\ln\left(\frac{1+\Pi}{1-\Pi} \right) ,\\
f_-(\Pi)&=-\frac{\Pi^3}{3}-\Pi+\frac{1}{2}\ln\left(\frac{1+\Pi}{1-\Pi} \right).
\end{align}
We will restrict our attention to positive energy modes, $\omega>0$. Next we construct wave packets on momentum space via
\begin{equation}
\Psi(\bar{T},\Pi)=\int_{0}^{\infty}d\omega~A(\omega)\,\psi^\omega(\bar{T},\Pi)
\end{equation}
where the most convenient choice for $A(\omega)$ is
\begin{equation}
A(\omega)=2 \sqrt{\frac{\beta ^{\alpha }}{\Gamma (\alpha )}}\, \omega ^{\alpha }\, \exp \left(-\frac{\beta}{2} \omega ^2\right),
\end{equation}
with parameters $\alpha,\beta>0$. $A(\omega)$ is centered around the energy (squared) $\omega^2=\frac{\alpha}{\beta}$ with width $\Delta\omega^2=\frac{\sqrt{\alpha}}{\beta}$. 

\begin{figure}
	\centering
	\begin{subfigure}{\columnwidth}
		\centering
		\includegraphics[width=\textwidth]{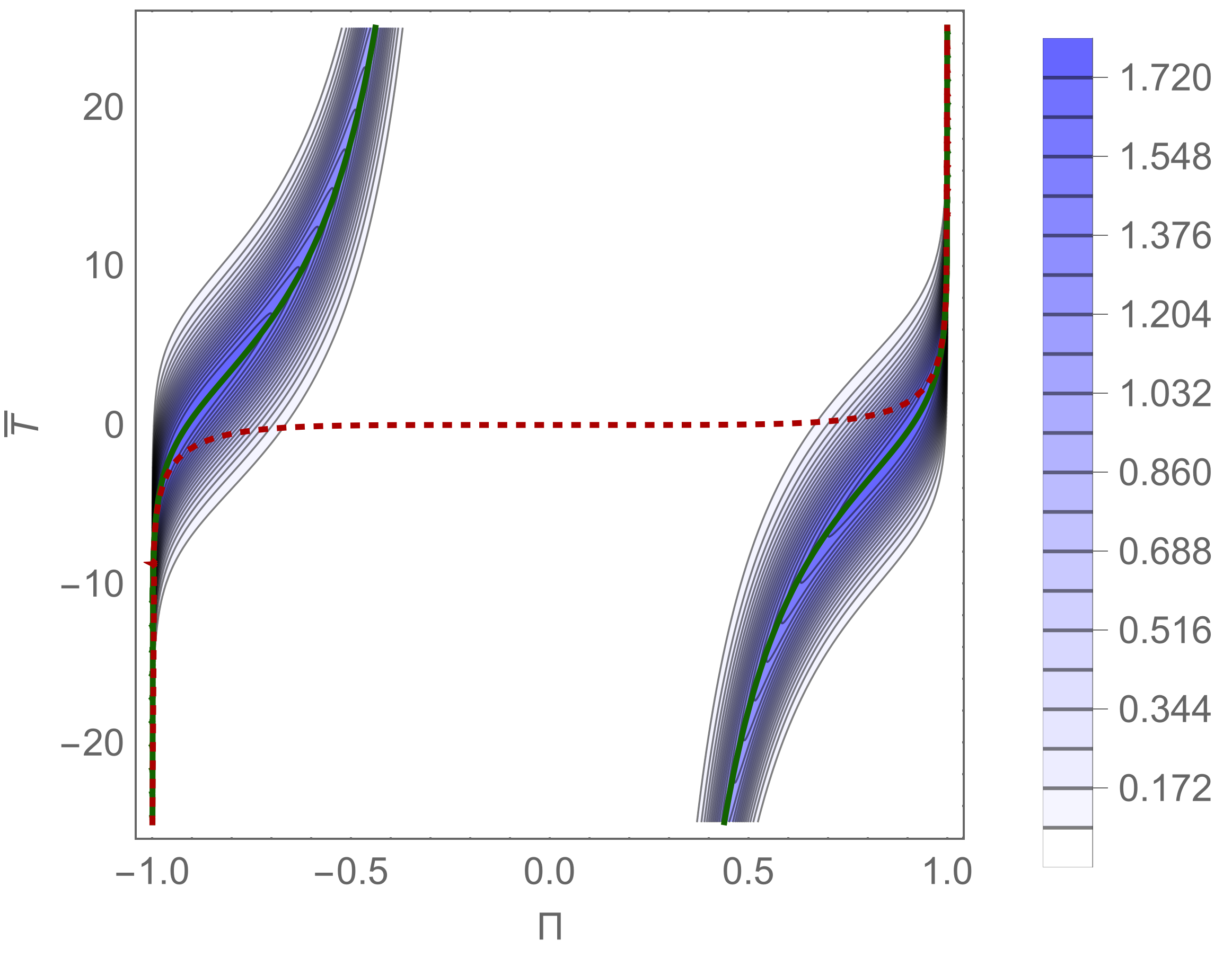}
		\caption{$C_+=1$ and $C_-=0$} \label{fig:wpa}
	\end{subfigure}
	\begin{subfigure}{\columnwidth}
		\centering
		\includegraphics[width=\textwidth]{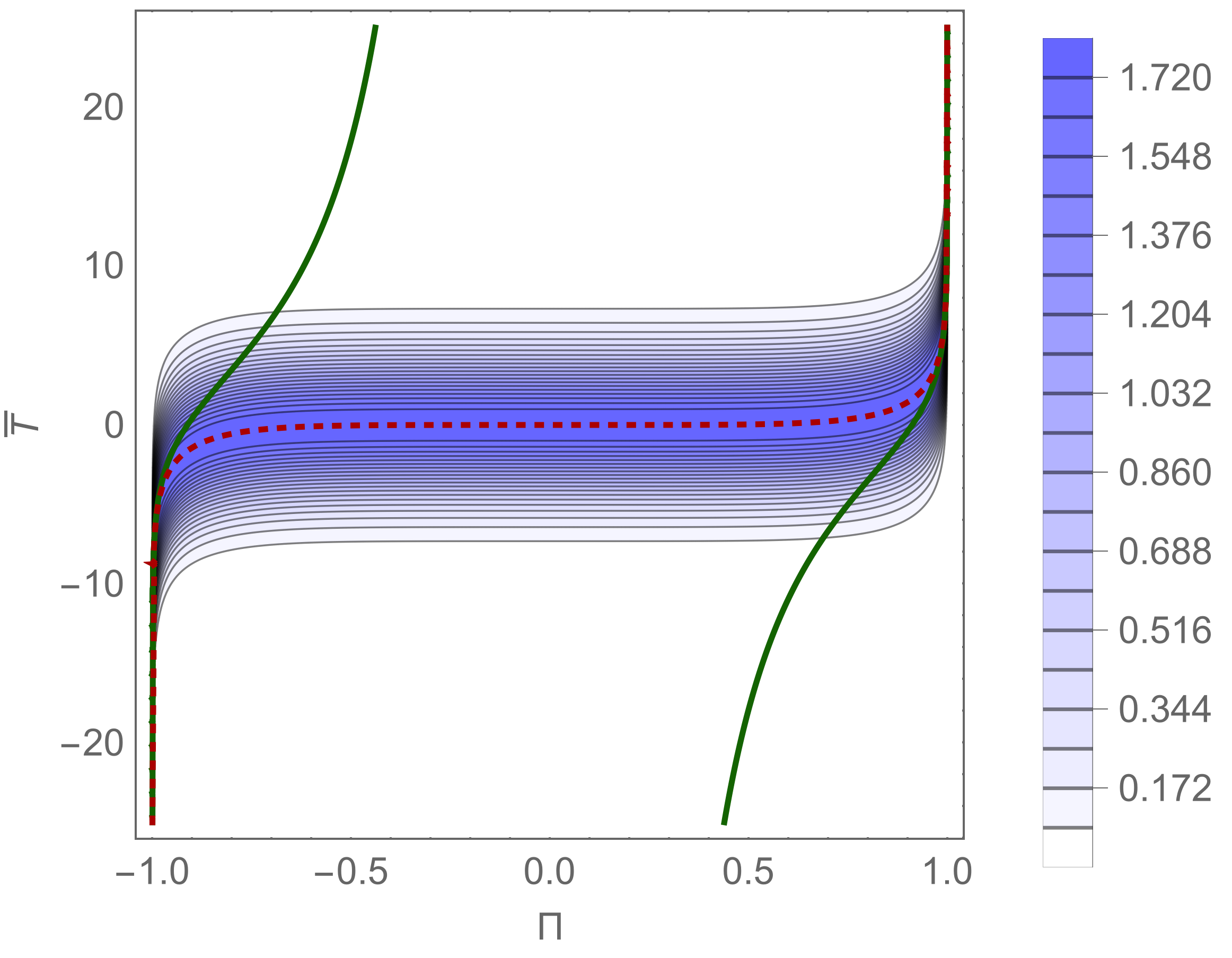}
		\caption{$C_+=0$ and $C_-=1$}
	\end{subfigure}
	\begin{subfigure}{\columnwidth}
		\centering
		\includegraphics[width=\textwidth]{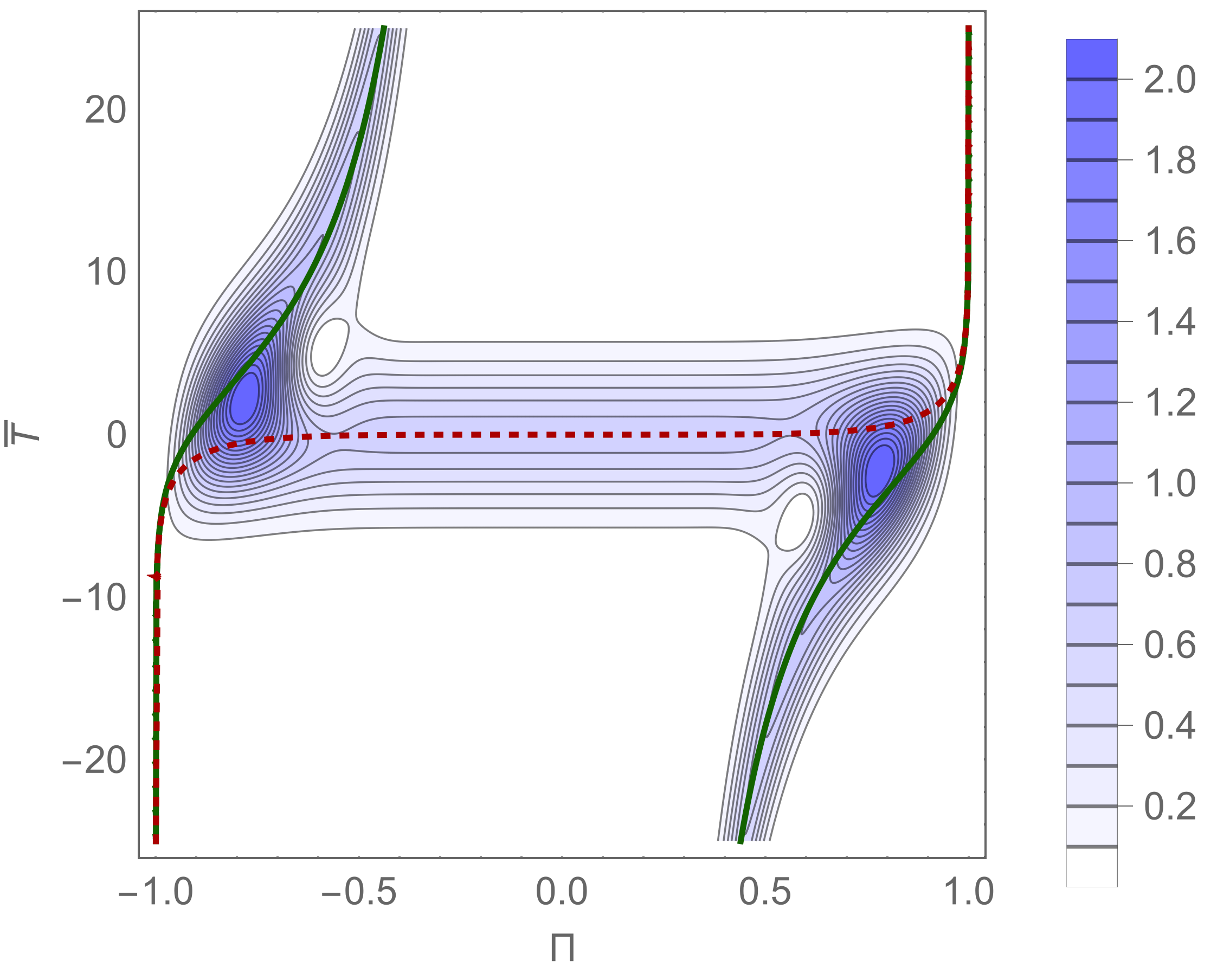}
		\caption{$C_+=0.725$ and $C_-=-0.595$} \label{fig:wpc}
	\end{subfigure}
	\caption{$|\Psi(\bar{T},\Pi)|^2$ for different contributions from $\bar{\mathsf{R}}>2\bar{P}_T$ and $\bar{\mathsf{R}}<2\bar{P}_T$ compared to the classical trajectories outside the horizon (full green lines) and inside (dotted red line) with energy $\bar{P}_T=\sqrt{\alpha/\beta}$, for $\alpha=10.1$ and $\beta=8.54$, in units where $\hbar=1$.}
	\label{fig:wavepacket_simple}
\end{figure}

The wave packet is plotted in Fig.\ \ref{fig:wavepacket_simple}. It can be given explicitly in terms of hypergeometric functions,
\begin{equation}
\Psi(\bar{T},\Pi)=C_+\Psi_+(\bar{T},\Pi)+C_-\Psi_-(\bar{T},\Pi),
\end{equation}
with
\begin{multline}
\Psi_\pm(\bar{T},\Pi)=\frac{2^{\frac{\alpha}{2}} \hbar ^{\frac{\alpha}{2}} \sqrt{\frac{\beta ^{\alpha }}{\Gamma (\alpha )}}}{\left( \beta  \hbar +4i\,f_\pm(\Pi)\right)^{\frac{\alpha+1 }{2}}}\left[\frac{2i \bar{T} \,\Gamma\! \left(\frac{\alpha }{2}+1\right) }{\sqrt{ \beta  \hbar +4i\,f_\pm(\Pi)}}\right.\\
\left.\times   \, _1F_1\!\left(\frac{\alpha +2}{2};\frac{3}{2};-\frac{1}{2\hbar}\frac{ \bar{T}^2}{\beta  \hbar +4 i\,f_\pm(\Pi)}\right)\right.\\
\left.+  \sqrt{2\hbar }\, \Gamma\! \left(\tfrac{\alpha +1}{2}\right)\!{}_1F_1\!\left(\frac{\alpha +1}{2};\frac{1}{2};-\frac{1}{2\hbar}\frac{ \bar{T}^2}{ \beta  \hbar +4 i\,f_\pm(\Pi)}\right)\right]\!.
\end{multline}

To interpret the results let us consider the classical equations of motion. Note first that
\begin{equation}
\frac{d \bar{\mathsf{R}}}{d\bar{T}}=\begin{cases}
-\frac{\sinh\frac{\bar{P}_\mathsf{R}}{\bar{\mathsf{R}}}}{\cosh^3\frac{\bar{P}_\mathsf{R}}{\bar{\mathsf{R}}}},~&\bar{\mathsf{R}}>2\bar{P}_T\\
\frac{\cosh\frac{\bar{P}_\mathsf{R}}{\bar{\mathsf{R}}}}{\sinh^3\frac{\bar{P}_\mathsf{R}}{\bar{\mathsf{R}}}},~&\bar{\mathsf{R}}<2\bar{P}_T
\end{cases},
\end{equation} 
from which we can read off that outside of the horizon, positive momentum $\bar{P}_\mathsf{R}$ corresponds to collapse toward the horizon, and negative momentum to expansion away from it. Inside of the horizon the situation is reversed, the sign of $\frac{d\bar{\mathsf{R}}}{d\bar{T}}$ always matching that of $\bar{P}_\mathsf{R}$, such that both inside and outside positive momentum means motion towards the horizon. This carries over to our new momentum $\Pi$. 

\begin{figure}
	\centering
	\begin{subfigure}{0.45\textwidth}
		\centering
		\includegraphics[width=\textwidth]{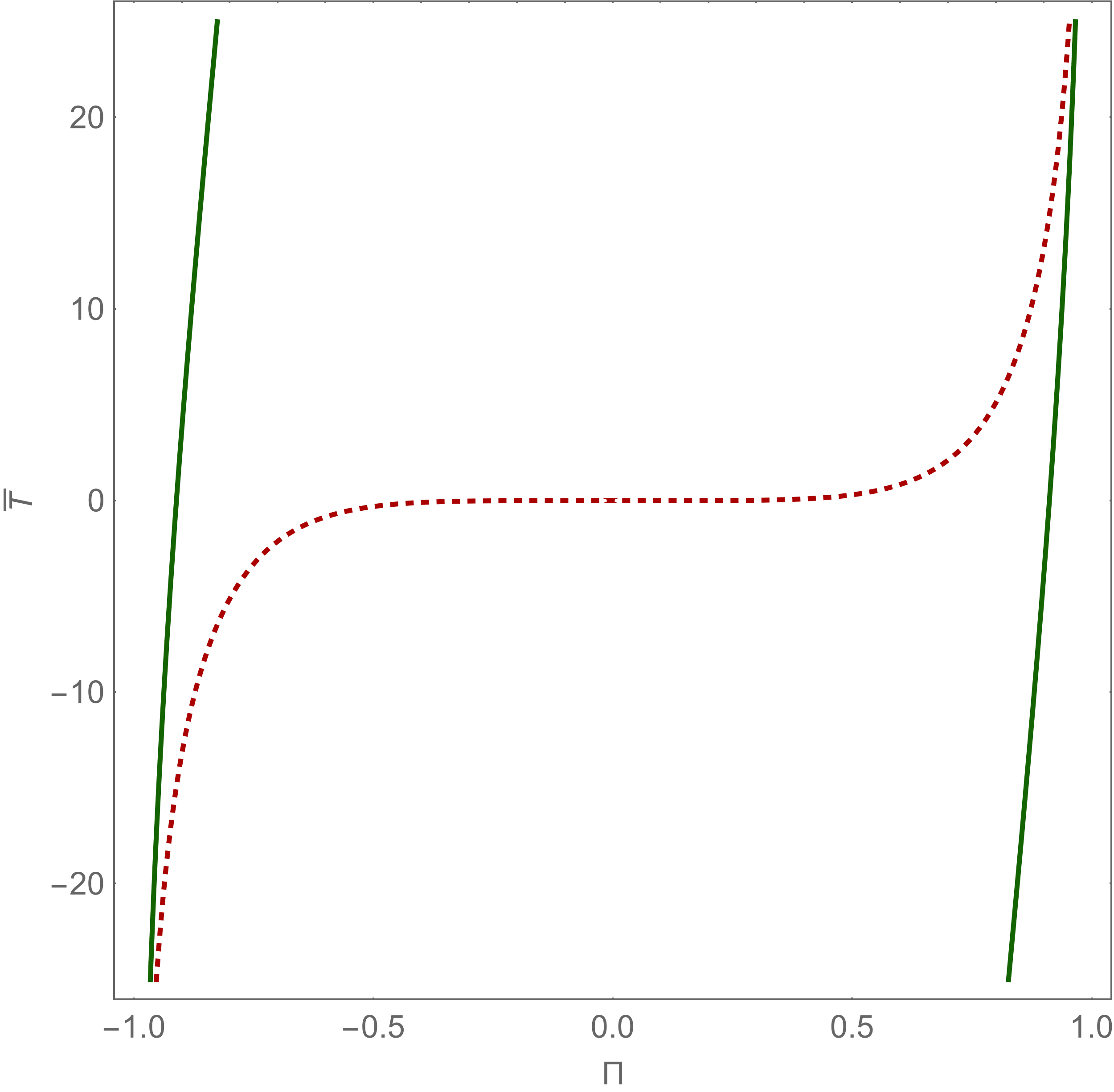}
		\caption{$\Pi(\bar{T})$ for $\bar{\mathsf{R}}>2\bar{P}_T$ (full green lines) and for $\bar{\mathsf{R}}<2\bar{P}_T$ (dotted red line).\\\hfill}
	\end{subfigure}
	\begin{subfigure}{0.45\textwidth}
		\centering
		\includegraphics[width=\textwidth]{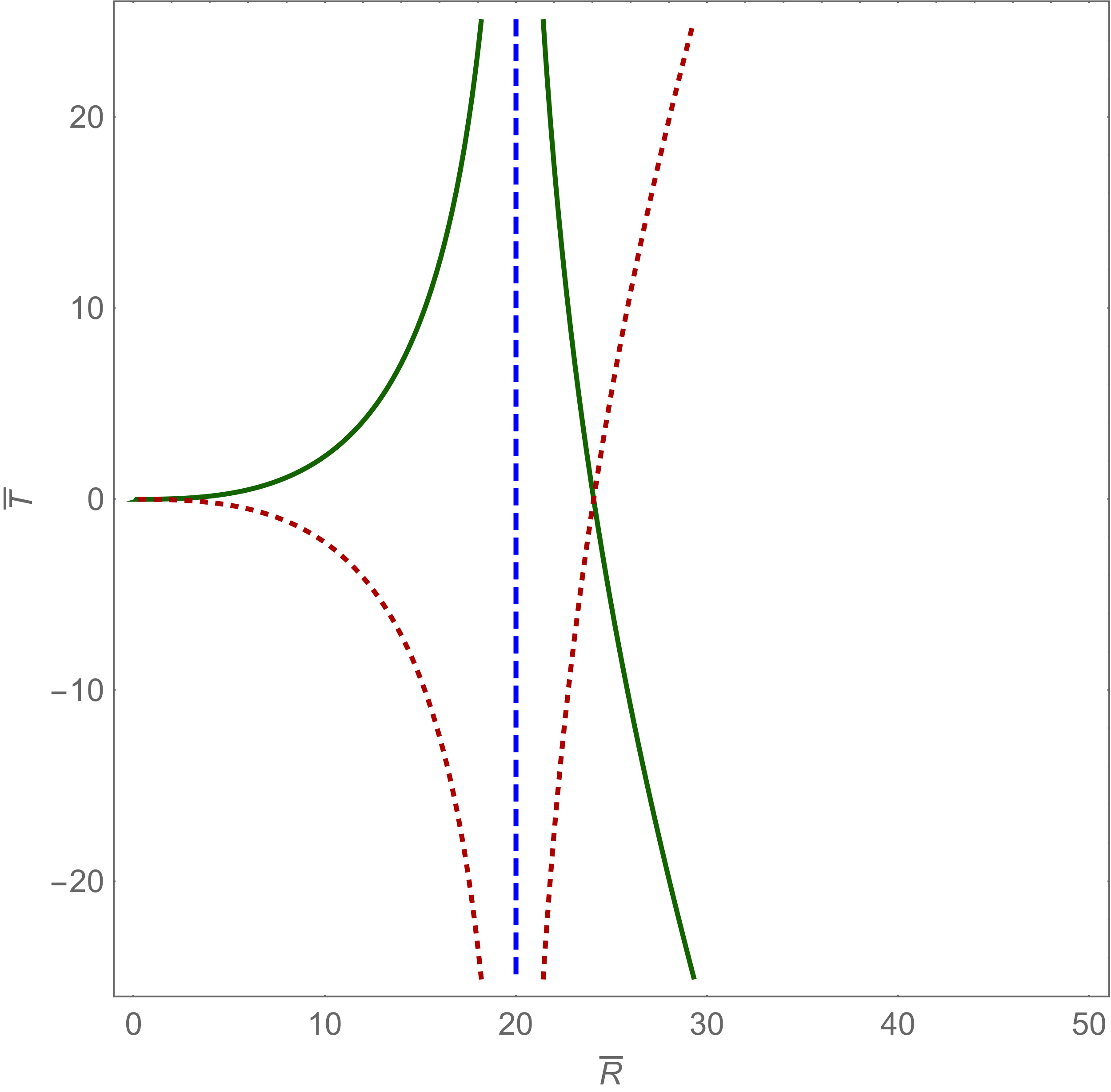}
		\caption{$\bar{\mathsf{R}}(\bar{T})$ for $\Pi<0$ (dotted red lines) and for $\Pi>0$ (full green lines), inside and outside the horizon (dashed blue line).}
	\end{subfigure}
	\caption{The classical trajectories $\Pi(\bar{T})$ and $\bar{\mathsf{R}}(\bar{T})$ inside and outside of the horizon for $\bar{P}_T=10$, in units where $\hbar=1$.}
	\label{fig:classical_trajectories}
\end{figure}

Furthermore we can see from \eqref{eq:hamiltonian_interior2} that near the horizon $\bar{P}_\mathsf{R}$ has to diverge, meaning $\Pi$ goes to $\pm1$, while far away from the horizon the momentum goes to $0$. Solving the equations of motion makes this more precise. For $\bar{\mathsf{R}}>2\bar{P}_T$, $\Pi(\bar{T})$ is given implicitly by
\begin{equation}
-\frac{1}{3\Pi^3}-\frac{1}{\Pi}+\frac{1}{2}\ln\left(\frac{1+\Pi}{1-\Pi} \right)=\frac{\bar{T}}{4\bar{P}_T},
\end{equation}
where $\bar{P}_T$ is a constant of motion. The trajectory in position space then follows from the above as $\bar{\mathsf{R}}=\frac{2\bar{P}_T}{\Pi^2}$. Analogously we compute for $\bar{\mathsf{R}}<2\bar{P}_T$
\begin{equation}
-\frac{\Pi^3}{3}-\Pi+\frac{1}{2}\ln\left(\frac{1+\Pi}{1-\Pi} \right)=\frac{\bar{T}}{4\bar{P}_T},
\end{equation}
and $\bar{\mathsf{R}}=2\bar{P}_T\Pi^2$. These trajectories are plotted in Fig.\ \ref{fig:classical_trajectories}.

By comparison to the classical trajectories we see that the wave packets in Figs.\ \ref{fig:wpa} and \ref{fig:wpc} represent dust clouds that start far away from the horizon and collapse toward it. After a transition they start to expand away from the horizon, matching the behavior that we saw in the dust proper time case: the cloud bounces.

Let us take a closer look at the transition between collapse and expansion: when only considering the dynamics as generated by the part of the constraint corresponding to $\bar{\mathsf{R}}>2\bar{P}_T$, meaning we set $C_-=0$, there is no transition; the trajectory in momentum space consists of two disconnected pieces. Only if we allow a contribution from inside of the horizon, $C_-\neq0$, are those two pieces connected. This suggests that the bounce here is facilitated by interference between inside and outside solutions. One could say that the inside leaks out into the near horizon region, leading to quantum gravitational corrections there.

We want to emphasize that our investigation is only fit to give a first indication of quantum corrected OS collapse from the point of view of an exterior observer. How exactly the interior and exterior dynamics mix, and what happens at the transition between collapse and expansion especially with regard to the horizon cannot be investigated in sufficient detail here. We hope to gain insight into at least some of these question by a more thorough investigation of the quantum dynamics of \eqref{eq:hamiltonian_interior2} in future work.

%%%%%%%%%%%%%%%%%%%%%%%%%%%%%%%%%%%%%%%%%%%%%%%%%%%%%%%%%%%%%%%%%%%%%%%%%
\section{Conclusions} \label{sec:chapter4}

In this paper we have laid the groundwork for a quantum OS model. First we have presented a consistent canonical formulation of the flat OS model, based on a partial symmetry reduction of spherically symmetric gravity with discontinuous dust, homogeneous in one region and vanishing outside of it, as the matter source. A particularly interesting feature was the emergence of the coordinate transformation between Schwarzschild Killing time and Painlev\'{e}-Gullstrand time from the canonical formalism.

Using Brown-Kucha\v{r} dust gave us access to the dust proper time as a canonical variable, and brought the Hamiltonian constraint into deparameterizable form with regard to this time. By performing the deparameterization when quantizing the system in this form we effectively took the point of view of the comoving observer.

This quantum theory is identical to the one from our treatment of the Lema\^itre-Tolman-Bondi model in \cite{MeLTB}, exhibiting singularity avoidance by a bounce. The comparison between inhomogeneous collapse in \cite{MeLTB} and homogeneous collapse here has revealed an interesting possibility: in \cite{MeLTB} we found indications that shell crossing becomes unavoidable close to the bounce, a feature which cannot show up in homogeneous models. Using the OS model it may thus be possible to further investigate instabilities of the homogeneous sector of a less symmetry-reduced quantum gravity model.

Back at the classical level we have then implemented Schwarzschild Killing time into the canonical formalism by promoting the aforementioned coordinate transformation to a canonical one. This gave us access to the Hamiltonian constraint in a form almost deparameterizable with regard to Killing time, and with that the point of view of the stationary observer.

Due to the unusual structure of the Hamiltonian constraint in this form we presented here a preliminary, heuristic quantization of it. The results nevertheless serve as a consistency check for our approach, since a bounce emerged as well; our method of switching observers classically can produce consistent quantum theories. Furthermore we have seen that the bounce is only present if one takes into account the inside of the horizon, pointing to a leaking out of this region into the outside as a possible mechanism for quantum gravitational corrections at the horizon.

The dynamics of the inside are of course somewhat arbitrary, since the relevant classical observer does not have access to this region. The prescription we used here did however emerge quite naturally from the canonical formalism.

In the future we plan to quantize the Hamiltonian constraint for the stationary observer in a more rigorous way, and will then be able to gain more insight into some of the open questions concerning the behavior of the horizon, the black hole lifetime, and how exactly the inside and outside mix, complementing our discussion in \cite{MeLTB}. This will hopefully lead us to a more complete picture of bouncing collapse.

Interesting possibilities for further work also are the inclusion of Hawking radiation, and the investigation of white hole instabilities. Both could influence the bouncing collapse scenario significantly, depending on the timescales involved.

%%%%%%%%%%%%%%%%%%%%%%%%%%%%%%%%%%%%%%%%%%%%%%%%%%%%%%%%%

\section*{Acknowledgments}

The author would like to thank Claus Kiefer, W{\l}odzimierz Piechocki, Nick Kwidzinski, Daniele Malafarina and Branislav Nikolic for helpful comments and discussions.

%\appendix
%\section{title}\label{app:A}

%%%%%%%%%%%%%%%%%%%%%%%%%%%%%%%%%%%%%%%%%%%%%%%%%%%%%%%%%%%%%%%%%%

%bibtex, remove before submission
%\bibliography{os_references}
%

%%%%%%%%%%%%%%%%%%%%%%%%%%%%%%%%%%%%%%%%%%%%%%%%%%%%%%%%%%%%%%%%%

\end{document}